\documentclass[aps,prl,nobibnotes,twocolumn,superscriptaddress,bibliography]{revtex4-2}
\usepackage{amsfonts}
\usepackage{mathrsfs}
\usepackage{amsmath}
\usepackage{color}
\usepackage{natbib}
\usepackage{graphicx}
\usepackage{bm}
\usepackage{amssymb}
\usepackage{xspace}
\usepackage{epstopdf}
\usepackage{dcolumn}
\usepackage{multirow}
\usepackage[colorlinks=true, letterpaper=true, pdfstartview=FitV, linkcolor=blue, citecolor=blue, urlcolor=blue]{hyperref}
\usepackage{wrapfig}
\usepackage[titletoc]{appendix}

\makeatletter

\newcommand{\Rmnum}[1]{\expandafter\@slowromancap\romannumeral #1@}
\makeatother

\begin{document}
\title{Topological Interstitial-Electron Conductor}

\author{Tingli He}
\affiliation{Key Lab of Advanced Optoelectronic Quantum Architecture and Measurement (MOE), Beijing Key Laboratory of Quantum Matter State Control and Ultra-Precision Measurement Technology, and School of Physics, Beijing Institute of Technology, Beijing 100081, China}
\affiliation{Anhui Provincial Key Laboratory of Magnetic Functional Materials and Devices, School of Materials Science and Engineering, Anhui University, Hefei 230601, China}

\author{Xiaoming Zhang}
\email{zhangxiaoming87@hebut.edu.cn}
\affiliation{State Key Laboratory of Reliability and Intelligence of Electrical Equipment, School of Materials Science and Engineering,
Hebei University of Technology, Tianjin 300401, China}

\author{Chaoxi Cui}
\affiliation{Key Lab of Advanced Optoelectronic Quantum Architecture and Measurement (MOE), Beijing Key Laboratory of Quantum Matter State Control and Ultra-Precision Measurement Technology, and School of Physics, Beijing Institute of Technology, Beijing 100081, China}

\author{Yilin Han}
\affiliation{Key Lab of Advanced Optoelectronic Quantum Architecture and Measurement (MOE), Beijing Key Laboratory of Quantum Matter State Control and Ultra-Precision Measurement Technology, and School of Physics, Beijing Institute of Technology, Beijing 100081, China}

\author{Yang Wang}
\affiliation{Key Lab of Advanced Optoelectronic Quantum Architecture and Measurement (MOE), Beijing Key Laboratory of Quantum Matter State Control and Ultra-Precision Measurement Technology, and School of Physics, Beijing Institute of Technology, Beijing 100081, China}

\author{Wei Jiang}
\affiliation{Key Lab of Advanced Optoelectronic Quantum Architecture and Measurement (MOE), Beijing Key Laboratory of Quantum Matter State Control and Ultra-Precision Measurement Technology, and School of Physics, Beijing Institute of Technology, Beijing 100081, China}

\author{Zhi-Ming Yu}
\email{zhiming\_yu@bit.edu.cn}
\affiliation{Key Lab of Advanced Optoelectronic Quantum Architecture and Measurement (MOE), Beijing Key Laboratory of Quantum Matter State Control and Ultra-Precision Measurement Technology, and School of Physics, Beijing Institute of Technology, Beijing 100081, China}

\author{Yugui Yao}
\email{ygyao@bit.edu.cn}
\affiliation{Key Lab of Advanced Optoelectronic Quantum Architecture and Measurement (MOE), Beijing Key Laboratory of Quantum Matter State Control and Ultra-Precision Measurement Technology, and School of Physics, Beijing Institute of Technology, Beijing 100081, China}

\begin{abstract}
Electron transport in solids arises primarily from two  mechanisms: freely moving  bulk electrons in metals, and gapless boundary states in topological insulators. Here, we report a new mechanism discovered in electrides. The topological interstitial-electron conductors (TIECs) proposed here are   insulating electrides, but host interstitial electrons (IEs) distributed within  crystal voids that traverse the entire unit cell. Without being tightly bound  to real ions, the IEs generally  experience low periodic potential barrier along the void channels. As a consequence, by applying a weak electric field sufficient to overcome the IE barriers but  far below the system's dielectric breakdown threshold, one can expect that the TIECs would generate a persistent current contributed by the IEs and propagating along the void channels.
We identify a family of realistic altermagnetic electrides, $A_5X_3$ ($A$ = Ca, Sr, Ba, Yb; $X$ = As, Sb), as  TIECs. Remarkably, for $A_5X_3$ materials,  the periodic potential barrier of the IEs along the void channels are ultralow, ranging from 13.43 to 67.96 meV per formula unit.
This  renders our proposal readily accessible to experimental verification.
We  further demonstrate  that when the IEs of $A_5X_3$ undergo periodic motion along the  channels, topological  surface states will emerge at the boundary perpendicular to the channel direction, and continuously move across the bulk band gap. This pumping-like behaviour not only corroborates the topological nature of TIECs, but also  rationalizes the finite-electric-field induced electronic transport within the  band theory. 
Our findings expand the classification  of electronic conductors, uncover  unexplored transport properties of electrides, and establish a  new material platform for low-power electronic devices.
\end{abstract}
\maketitle

\begin{figure*}[tb]
	\includegraphics[width=16 cm]{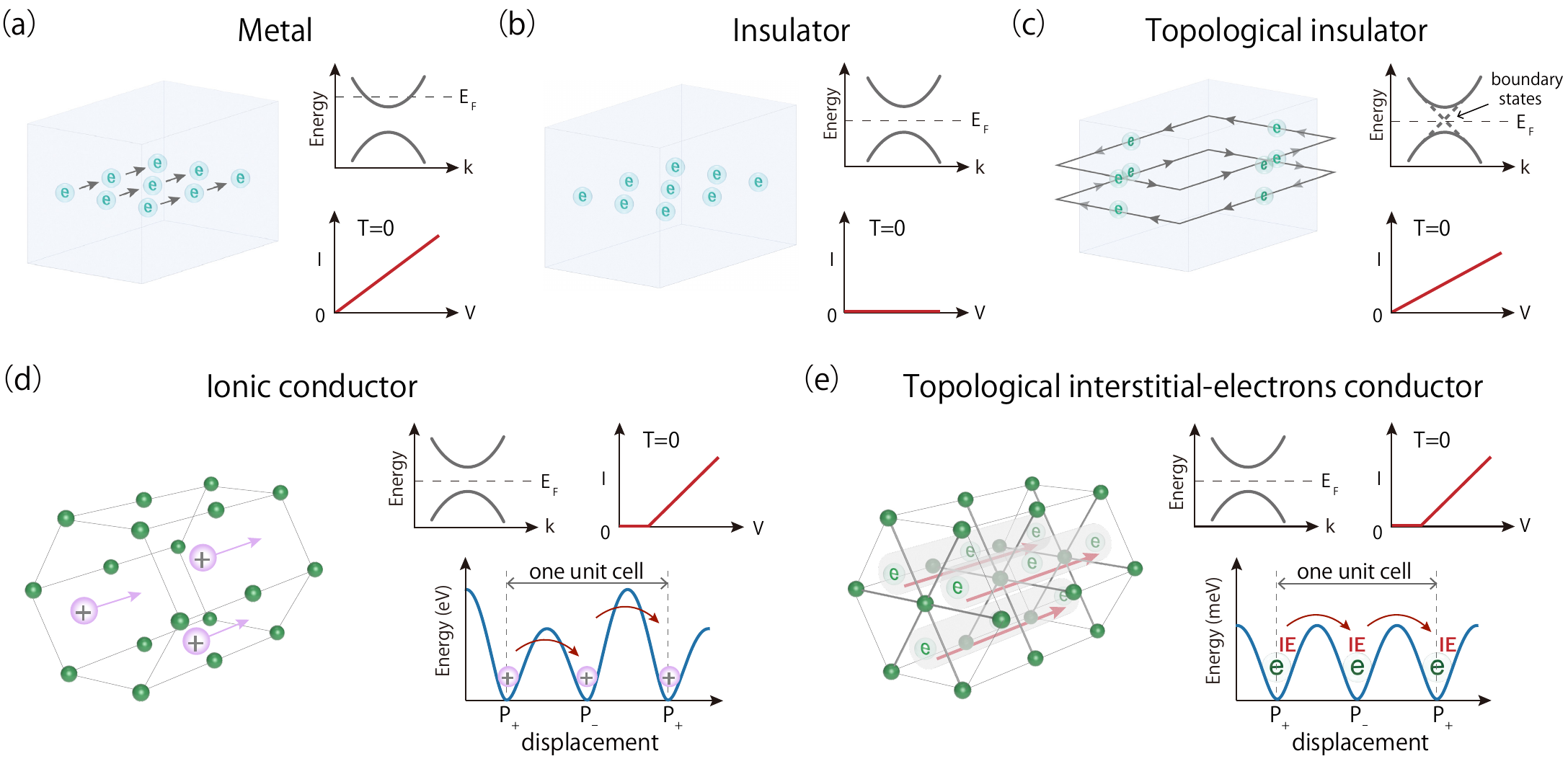}
	\caption{Schematic of the charge transport contributed by (a,c) freely moving electrons, (d) ``trapped" ions and (e) ``trapped" interstitial electrons. 
(a) Conventional metal and (b) insulators host  partially and fully occupied  bands, leading to finite and negligible conductivity at zero temperature ($T=0$). (c) Topological insulator, which host several conductive channels from  gapless boundary states.
The insets in (a-c) illustrate the  band structure and $I$-$V$ curve of the  state.
(d) Ionic conductor  is a band insulator but has ion-conducting channel. To conduct charge, the applied electric field should overcome the energy barrier of ions under displacement.
(e) Topological  interstitial-electrons conductor is also a band insulator but can conduct charge via the interstitial electrons. 
Since the interstitial electrons are lighter and more extension, they generally  have lower energy barrier under displacement.
The insets in (d-e) illustrate the  band structure, $I$-$V$ curve, and   energy barrier.}
	\label{fig1}
\end{figure*}

In the framework of band theory, solid-state systems are classified into metals and insulators based on their  band occupation. Metals possess partially filled electronic bands, allowing delocalized bulk electrons to move freely under an infinitesimal electric field [see Fig. \ref{fig1}(a)]. In contrast, insulators have fully occupied valence bands and empty conduction bands separated by a finite band gap, resulting in negligible conductivity at zero temperature [see Fig. \ref{fig1}(b)].
The topological insulators   expanded this classification \cite{BernevigHughes+2013,shen_topological_2017}, which  are  insulators with a bulk band gap, but  host gapless topological boundary states that enable conduction along  boundaries [see Fig. \ref{fig1}(c)].
Another intriguing mechanism of electron transport in insulators is topological pumping \cite{PhysRevB.27.6083}, where a periodic finite (not  infinitesimal) perturbation can pump electrons from one side to the other in an insulating system. But topological pumping is a dynamical process.
Beyond these electronic conductors, there also exist ionic conductors \cite{Ohno_2020,Wujpclett,WuPRL}, which are band insulators but support charge transport via ions--a process that is, however, associated  with mass transfer, as illustrated in Fig \ref{fig1}(d).
This raises a fundamental question: \emph{can a system exist that is intrinsically insulating in both  bulk and  boundary, yet still conducts electrons under a static electric field?}

In this work, we answer this  paradoxical question in the affirmative by identifying a new transport mechanism in electrides.
Electrides are a unique class of crystalline materials \cite{dye1993anionic, dye2009electrides,hosono_advances_2021, doi:10.1021/jacs.3c00284, Shenacs.accounts.4c00394}, characterized by the presence of interstitial electrons (IEs) that act as anions confined in the interstitial voids of the crystal lattice.
Due to the dual nature of localization and extension of IEs, electrides exhibit many  intriguing properties, and have significant application potential in various  fields, such as electron emission \cite{huang1990low, 10.1063/1.2149989}, spintronic devices \cite{doi:10.1021/acsanm.5c01074, doi:10.1021/acsanm.5c01074}, ammonia synthesis \cite{rafiqul2005energy, kitano2016essential,https://doi.org/10.1002/anie.201712398, wu_intermetallic_2019}, and high-performance catalysts \cite{LI201445,khan2018facile, mengAS}.
Various materials are predicted as  electride, and some of them  have been experimentally confirmed  \cite{HosonoChemicalReviews, DruffelJACS,WangJunjieJACS, QiangZhengSA,HuangHuaqingNL}.
Interestingly, the interstitial voids in many electrides are not isolated but interconnected, forming one-dimensional (1D) void channels or even 2D void planes. 
Moreover, since the IEs are lighter and more extension than real ions, they generally have lower energy barrier under displacement.
All these observations indicate that even for the electrides that are band insulators, these connected voids can serve as unconventional  transport channels for the ``trapped" IEs [see Fig. \ref{fig1}(e)], resulting in persist charge transport under finite electric field, similar to the case in ionic conductors.

To demonstrate our idea, we  study the  displacement of the IEs in  electride material family  $A_5X_3$ ($A$ = Ca, Sr, Ba, Yb; $X$ = As, Sb), which can be captured by the  ferroelectric (FE)  properties contributed by the  IEs of the systems.
All the $A_5X_3$  materials share similar crystal structure and are altermagnetic (AM) systems.
Particularly, these materials feature 1D  void channels that traverse the entire unit cell along  $c$ direction, in which two spin-polarized IEs are localized.
We find that the $A_5X_3$'s  are  type-II fractional quantum ferroelectricity (FQFE) with ultralow  potential barrier.
Importantly, the two fractional FE states of $A_5X_3$ can be  switched by \emph{solely }shifting the  two IEs along the  void channel by half-integer lattice constant, without requiring any displacement of the  real ions.
Moreover,  in $A_5X_3$,  a subsequent translation of the IEs by an additional half-integer lattice constant along the channel returns the system to its initial state, up to an integer lattice translation of the IEs along the $c$-axis, as  illustrated in Fig. \ref{fig1}(d).
All these results  directly imply that  a finite static electric field can generate a persistent direct current in $A_5X_3$, similar to  the  case in  ionic conductors but without ion transfer.

Furthermore, we find that  as the IEs undergo periodic translation along the void channel, a  surface state emerges at the $(001)$ surface of $A_5X_3$, and  continuously move across the  bulk band gap.
By treating the periodic  evolution  of the IEs  as an additional   dimension, the 3D $A_5X_3$ system can be mapped onto a $(3+1)$D topological system with a nontrivial Chern number ${\cal{C}}=2$,  analogous to the topological pumping process.
Therefore, we denote this novel phenomenon  as topological IE transport, and classify the systems hosting such  behavior as TIECs.

\global\long\def\arraystretch{1.2}%
\begin{table*}[tb]
	\caption{Magnetic, ferroelectric,  and electride properties of  $A_5X_3$ materials.}
	\centering
	\begin{ruledtabular}
		\begin{tabular}{@{}lclccccc@{}}
			 & AM order  & Band gap (eV)& Switching barrier & Polarization & $Q_c$ & ELF & Experimental \\
			&  ${\cal M}$ ($\mu_B$) & (GGA/HSE/exp.) &  (meV/f.u.) &  ($\mu$C/cm$^2$) & ($\mu$C/cm$^2$) & value & synthesis \\
			\hline
			Ba$_5$Sb$_3$ & 0.58 & 0.30/0.56/0.30 \cite{wolfe1990electrical} & 67.96 & 8.5  & 17 & 0.984 & Yes \\
			Sr$_5$Sb$_3$ & 0.51 & 0.26/0.57/-- & 39.09 & 9.5 &19 &0.971  & No \\
			Ca$_5$As$_3$ & 0.45 & 0.39/0.81/-- & 34.05 & 12.0 & 24  & 0.966 & No \\
			Ca$_5$Sb$_3$ & 0.43 & 0.11/0.43/-- & 23.20 & 10.5  & 21 & 0.950 & No \\
			Yb$_5$Sb$_3$ & 0.36 & 0.04/0.20/0.14 \cite{PhysRevB.98.125128} & 13.43 & 10.0  & 20 & 0.922 & Yes \\
		\end{tabular}
	\end{ruledtabular}
	\label{table1}
\end{table*}

\textit{\textcolor{blue}{Lattice  and  band structure of $A_5X_3$.—}}
Previous  works \cite{wolfe1990electrical,PhysRevB.98.125128, 10.1063/5.0187372} have shown that the $A_5X_3$ are  electrides with an AM  ground state, which is further confirmed by our calculations, as shown in  Supplemental Material (SM) \cite{SM}.
Particularly, Ba$_5$Sb$_3$ and Yb$_5$Sb$_3$ have been experimentally synthesized \cite{wolfe1990electrical,PhysRevB.98.125128}, and Yb$_5$Sb$_3$ has previously  been demonstrated as spin-splitting antiferromagnetic (AFM) material \cite{PhysRevB.98.125128}, which is exactly the AM material \cite{PhysRevX.12.031042}.
Furthermore, Yb$_5$Sb$_3$ electride  still maintains its AM semiconductor properties at room temperature in experiment \cite{PhysRevB.98.125128}.

\begin{figure}
	\includegraphics[width=0.95\columnwidth]{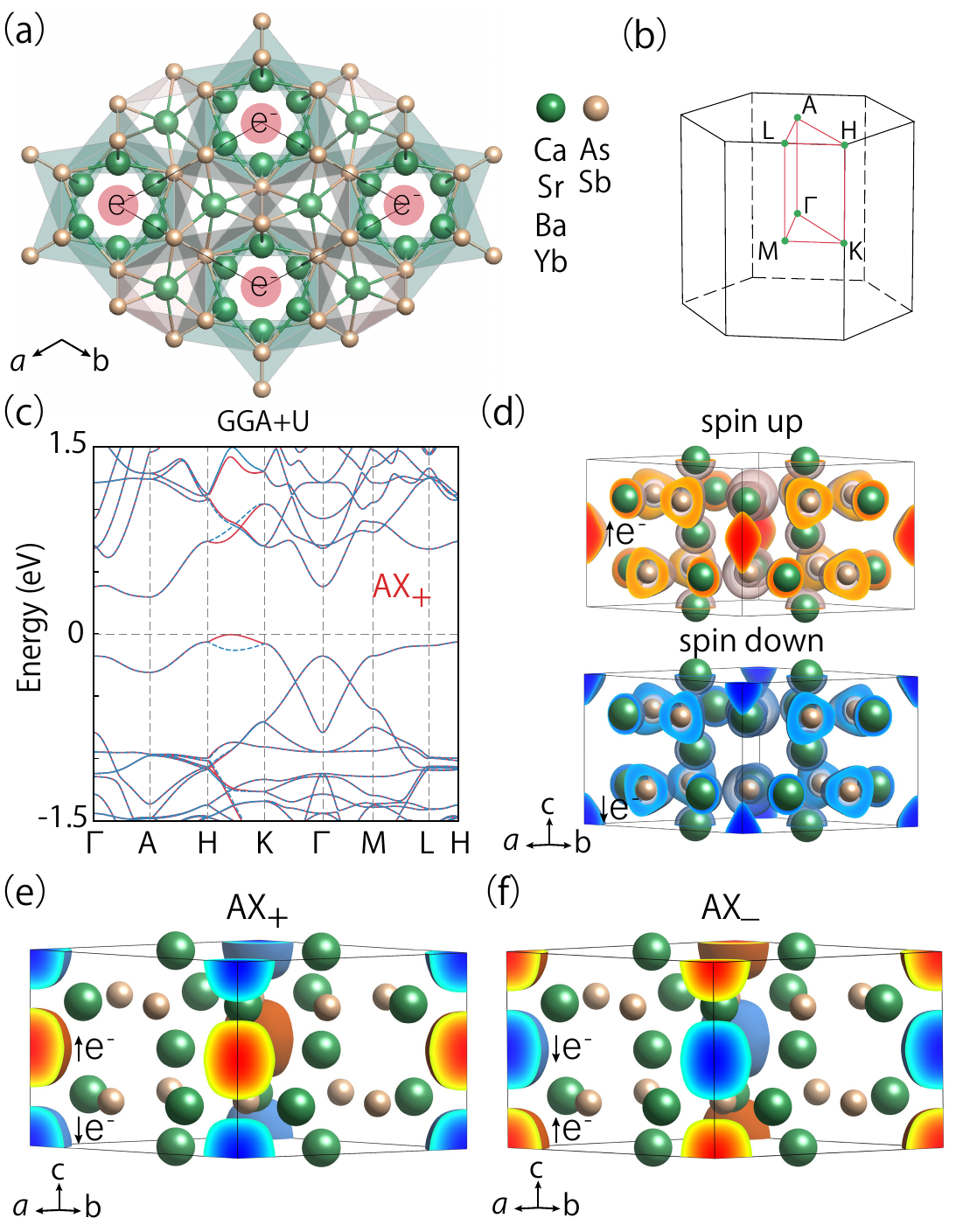}
	\caption{ (a) Crystal structure  and (b)  Brillouin zone (BZ) of $A_5X_3$  materials.
		 (c) The electronic band structure of Ba$_5$Sb$_3$  without  SOC calculated from   GGA+U  method. In (c), the red solid lines and blue dashed lines represent spin-up and spin-down bands, respectively. (d) ELF of spin-up and  spin-down channels for Ba$_5$Sb$_3$. (e-f) Two energy-degenerate ground states of Ba$_5$Sb$_3$: $AX_+$ and $AX_-$, which exhibit opposite AM order. $AX_+$ and $AX_-$ are connected by both ${\cal{T}}$ and $\{C_{2z}|00\frac{1}{2}\}$. 
		 Spin density distribution of Ba$_5$Sb$_3$ for (e) $AX_+$ and (f) $AX_-$.  }
		\label{fig2}
\end{figure}

The $A_5X_3$  materials crystallize in a  three-dimensional (3D) hexagonal crystal  with space group $P6_3/mcm$ (No. 193), as shown in Fig. \ref{fig2}(a).
The metal atoms $A$ occupy two inequivalent Wyckoff positions, namely $4d$ and $6g$, while the $X$ atoms reside at the $6g$ Wyckoff position.
A key structural feature of the $A_5X_3$ is  the presence of 1D straight  void channel  that traverse the entire unit cell along the $c$-direction [see  Fig. \ref{fig2}(a)].

We have employed both  generalized gradient approximations (GGA)+U method  and  hybrid functional approach  (HSE06) method to calculate the band structure of $A_5X_3$.
The two methods give qualitatively consistent band structures \cite{SM}, indicating  that the GGA+U method with a large  $U$ is acceptable for studying $A_5X_3$  materials. 
Therefore, unless otherwise specified, our following calculations will  use the GGA+U method.

The band structures of Ba$_5$Sb$_3$ without spin-orbit coupling (SOC) is plotted in Fig. \ref{fig2}(c), a spin splitting can be clearly observed along H-K path, which is  a typical feature of altermagnetism. The other $A_5X_3$  materials share similar band structures \cite{SM}. 

Importantly, both GGA+U and HSE methods  show all the $A_5X_3$   are   electride materials \cite{SM}, as evidenced by the spin-resolved electron localization function (ELF) of the system.
From  Fig. \ref{fig2}(d),  one can find that there indeed  exist  excess electrons localized in the  crystal voids.
Specifically, the IEs for the spin-up subspace  reside at the high-symmetry  position  $(00\frac{1}{2})$, while those  for the spin-down  subspace are localized  at the $(000)$ position.
Notably, the IEs in  $A_5X_3$  materials are  considerable  spatial extension along the $c$-direction, as shown in Figs. \ref{fig2} (d) and SM \cite{SM}.

The  electride feature of $A_5X_3$ can be more clearly observed by the spin density distribution [see Fig. \ref{fig2}(e)].
Remarkably, the  spin density distribution shows that the magnetism of $A_5X_3$ solely  originates from the IEs, which is in sharp  contrast to conventional magnetic materials that rely on real magnetic atoms.
The IEs  in $A_5X_3$ contributes a sizable  magnetic moment (see Table \ref{table1}), and are arranged in an antiparallel configuration.
Thus, the  net magnetic moment for $A_5X_3$ is zero.
Taking magnetism into account, the magnetic space group of $A_5X_3$  becomes $P6_{3}'/m'cm'$ (No. 193.259), which has spatial inversion ${\cal{P}}$ and a combined operation $\{C_{2z}|00\frac{1}{2}\}{\cal{T}}$ with $\{C_{2z}|00\frac{1}{2}\}$ and ${\cal{T}}$ being two-fold screw rotation and time-reversal symmetry, respectively.
In $A_5X_3$, the two spin-polarized  IEs  are connected by $\{C_{2z}|00\frac{1}{2}\}{\cal{T}}$ symmetry rather than ${\cal{PT}}$ or ${\cal{T}}t$ with $t$ being fractional translation, which further demonstrate that the $A_5X_3$ is an AM system \cite{PhysRevX.12.031042}. 

\textit{\textcolor{blue}{Screw-rotation connected ground states  and  type-II FQFE.—}}
According to the magnetic configuration, one easily knows that  the $A_5X_3$  has two energy-degenerate ground  states with opposite  AM order, as illustrated in Fig. \ref{fig2}(e-f), which here are labeled as $AX_+$ and $AX_-$. The electronic band structure of the two states have opposite spin-splitting \cite{SM}.
To quantitatively describe the  altermagnetism of  $A_5X_3$, we introduce an AM order, defined as
\begin{eqnarray}\label{eq:1}
{\cal M}=  \int_{z\in(c/4,3c/4)} m({\bm r}) d {\bm r}- \int_{z\in(-c/4,c/4)} m({\bm r})d {\bm r} ,
\end{eqnarray}
where  $m=(m_{\uparrow}-m_{\downarrow})/2$ with $m_{\uparrow}(m_{\downarrow})$  the local   density of spin-up (spin-down) electrons of the system, and the origin of coordinates in  Eq. (\ref{eq:1})  is chosen to be  ${\cal{P}}$-invariant.
Notice that  ${\cal M}$ in  Eq. (\ref{eq:1}) is not the conventional  N\'eel vector.
For $AX_+$ and $AX_-$ of Ba$_5$Sb$_3$ in Fig. \ref{fig2}(e-f),  the AM  order ${\cal M}$  are opposite and calculated to be $\pm 0.58~\mu_B$, as listed in Table \ref{table1}.

Generally, the  $AX_+$ can be obtained from the $AX_-$ by reversing the spin of the IEs, as the two states are connected by  ${\cal{T}}$ symmetry.
However, due to the presence of $\{C_{2z}|00\frac{1}{2}\}{\cal{T}}$ in $A_5X_3$, $AX_+$ and $AX_-$ can also be connected by spatial operator   $\{C_{2z}|00\frac{1}{2}\}$. 
This indicates that $AX_+$ and $AX_-$ can also be reversibly switched \emph{solely} through the movement of IEs along the 1D void channels, without involving any ion displacement.

\begin{figure*}
	\includegraphics[width=1.95\columnwidth]{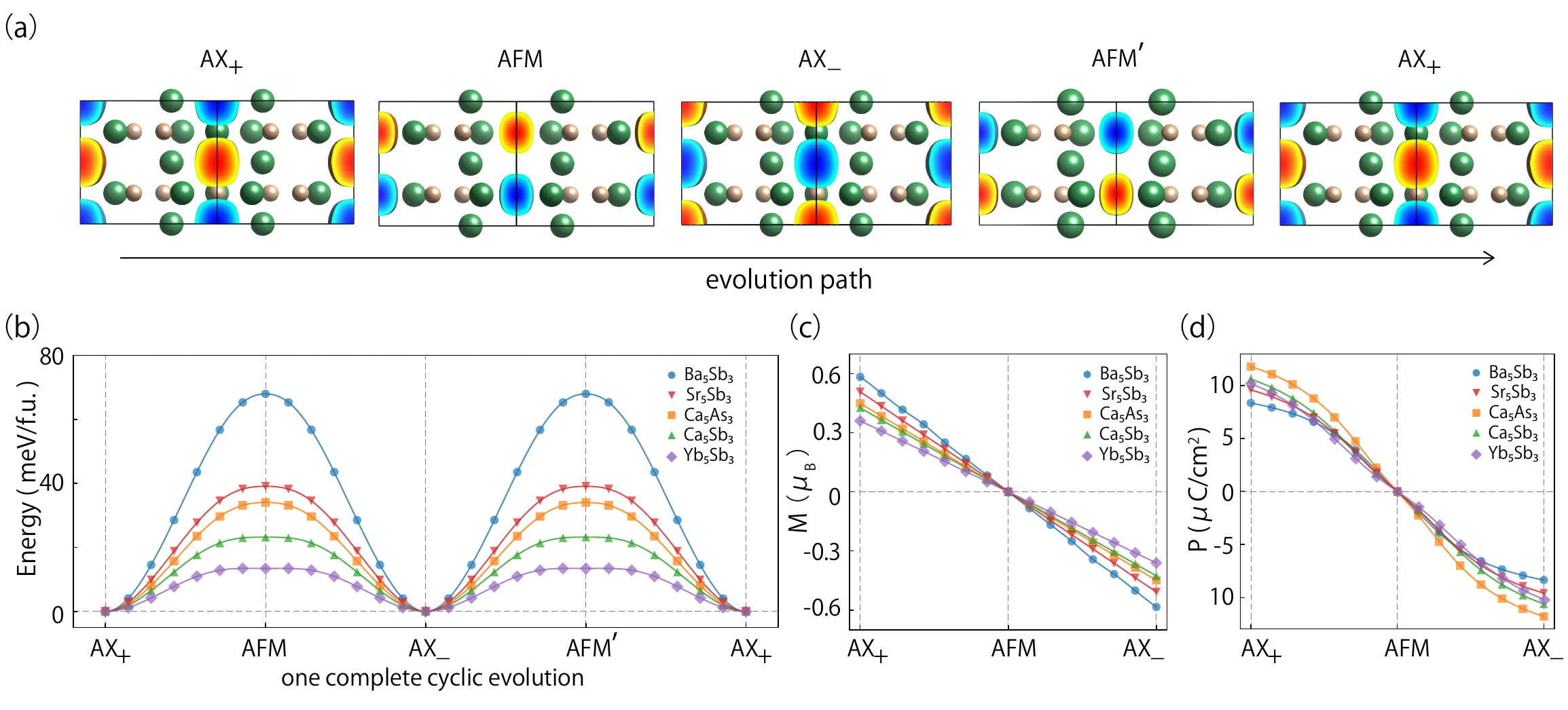}
	\caption{(a) Spin density   distribution of   $A_5X_3$  along the  $AX_+$--AFM--$AX_-$--AFM$'$--$AX_+$  evolution path. (b) The periodic potential barrier barrier of $A_5X_3$ along the   evolution path in (a).  (c) AM order  ${\cal M}$ and (d) electric polarization of  $A_5X_3$ along  $AX_+$--AFM--$AX_-$ path without SOC. The band structures of $A_5X_3$ along the path are shown in SM \cite{SM}.}
	\label{fig3}
\end{figure*}

To directly  demonstrate this unconventional  feature of $A_5X_3$, we calculate the energy barrier and electric polarization  along a direct path connecting $AX_+$ and $AX_-$.
Along this path [see Fig. \ref{fig3}(a)], the spin-up IEs move from $(00\frac{1}{2})$ to $(00\frac{3}{4})$ and finally to $(001)$ position, while the spin-down IEs evolve from $(000)$ to $(00\frac{1}{4})$  and then to $(00\frac{1}{2})$.
Interestingly, when the  two IEs respectively moving to  $(00\frac{1}{4})$ and $(00\frac{3}{4})$ positions, the system become a  high symmetric AFM phase with ${\cal{PT}}$ symmetry \cite{SM}.
The energy barriers along this path for $A_5X_3$ materials  range from 13.43 to 67.96 meV/f.u. [see Fig. \ref{fig3}(b) and Table \ref{table1}].
All the energy barrier are rather low, showing  that the IE switching of  all the $A_5X_3$ materials proposed here are accessible in the experiment.

The variation of ${\cal M}$ and the electric polarization of $A_5X_3$ along the evolution path are shown in Figs. \ref{fig3}(c-d). 
Along the path, ${\cal M}$ gradually decreases and vanishes at the  AFM state, but reappears--with its sign inverted--after passing through it [see Fig. \ref{fig3}(c)]. 
The evolution of the  AM spin splitting  of   $A_5X_3$ along the path is similar to  that of ${\cal M}$ \cite{SM}.
This directly shows that the AM order ${\cal M}$ defined in Eq. (\ref{eq:1}) is appropriate  for characterizing altermagnetism and AM spin splitting beyond the N\'eel vector.
Moreover, the  electric polarization  of all the $A_5X_3$ materials continuously changes along the path, and take a universal value at $AX_+$ ($AX_-$) state in the unit of polarization quantum $Q_c$ , i.e. the polarization  ${\bm P}=(00\frac{\pm Q_c}{2})$  (see Table \ref{table1}). Here, $Q_c = e R_c/V_\text{cell}$, $R_c$ is the $c$-direction lattice constant  and $V_\text{cell}$ is the unit cell volume.
Therefore, the total electric polarization difference for $A_5X_3$ materials  is an integer times of $Q_c$, leading to  a type-II FQFE \cite{PhysRevLett.134.016801,AMFQFE,luo2026FE}.

Clearly, the electric  polarization  of  $A_5X_3$ is only contributed by the electrons, especially the IEs, and has no  contribution from real ions, as  the two states $AX_+$ and $AX_-$ are identical in their atomic configurations.
This is a unique phenomenon due to the electride feature of $A_5X_3$  and has not yet  been found in any previously reported  materials.
Moreover,  the $A_5X_3$ is a novel multiferroic material featuring  type-II FQFE and altermagnetism. 
Due to the type-II FQFE characteristics, the polarization of all $A_5X_3$ materials is relatively large, with the polarization ranging from 17 to 24 $\mu$C/cm$^2$ (see Table \ref{table1}). This polarization is even comparable to that of  type-I multiferroic materials,  such as Bi(Yb)FeO$_3$ \cite{science.1080615,jeong2012}.

At last, we have  checked the influence of SOC on the switching barrier, the electric polarization and the  N\'eel vector of $A_5X_3$.
We find that the results are qualitatively the same  \cite{SM}.
This may result from the fact that the phenomena discussed here are solely or mainly induced by the IEs, which are not sensitive to the  SOC effect.

\textit{\textcolor{blue}{IE transport and its topological feature.—}}
The type-II FQFE feature of $A_5X_3$ directly suggests  a novel  electron transport that is  fundamentally distinct from conventional conduction paradigms. As aforementioned,  the two degenerate ground states $AX_+$ and $AX_-$ of $A_5X_3$ can be reversibly switched via a half-integer lattice translation of IEs along the 1D void channels without any displacement of the real ions. A remarkable consequence is that  a subsequent half-integer translation of IEs along the same direction returns the system to its initial magnetic configuration, i.e.  $AX_+$  state [see Fig. \ref{fig3}(a)], differing only by an integer lattice translation of the IEs along the $c$-axis. The two evolution paths from $AX_+$ to  $AX_-$ and from $AX_-$ to  another  $AX_+$ as a whole shows that the IEs  experience a ultralow periodic potential barrier  along the 1D void channels [see Figs.  \ref{fig3}(a-b)]. 
Notice that for all the  $A_5X_3$  materials, the potential barrier  is much smaller that the band gap of the system.

 \begin{figure}
 	\includegraphics[width=0.99\columnwidth]{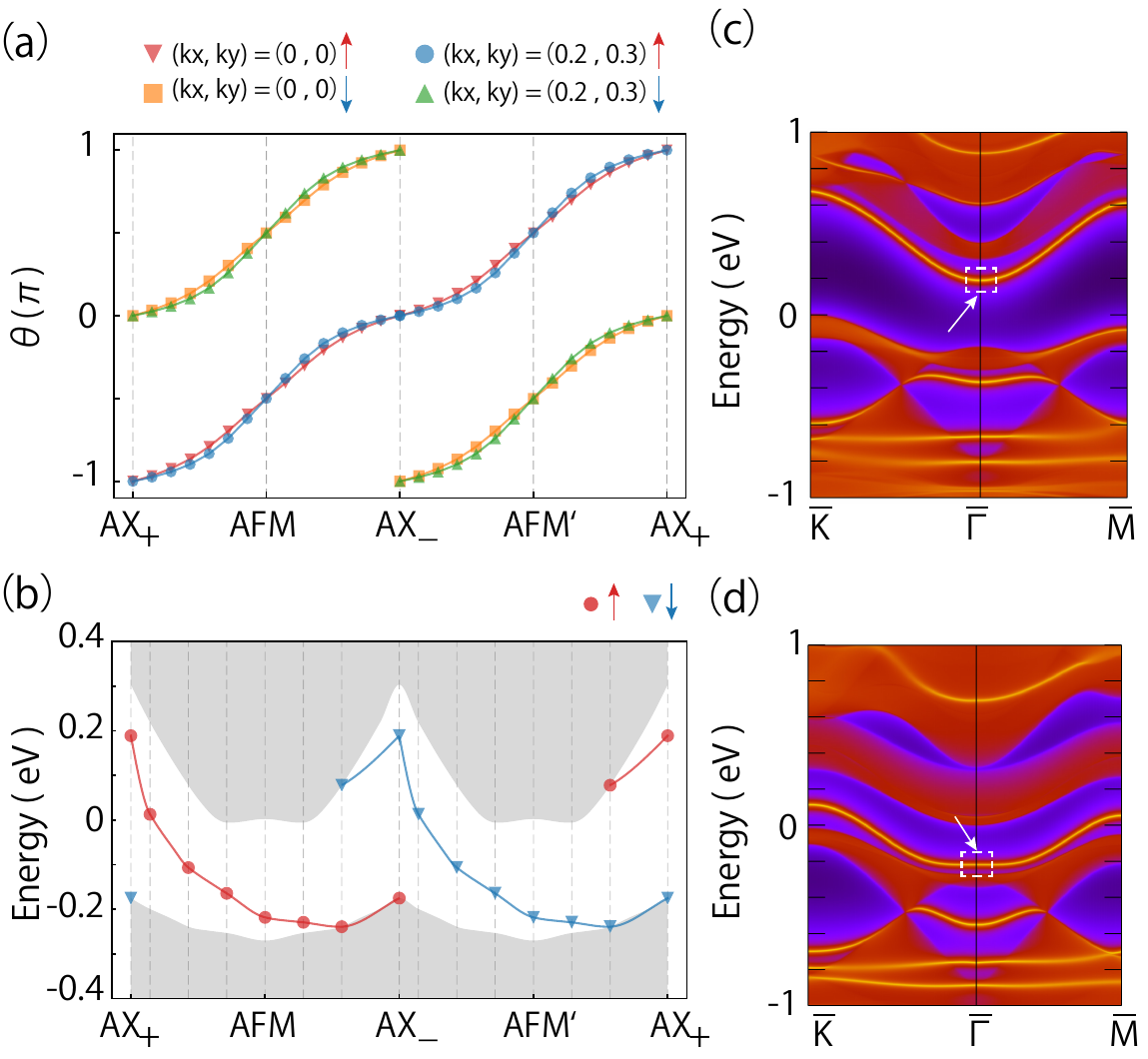}
 	\caption{(a) The $z$-direction  Wannier center of Ba$_5$Sb$_3$.
 		(b) The evolution of the energy of the (001) surface state at the $\bar{\Gamma}$ point, marked in (c) and (d). 
 The gray area denotes the bulk state at ${\Gamma}$ point along the evolution path.  (c-d) (001) surface states of Ba$_5$Sb$_3$  in the spin-up channel for  (c) the   $AX_+$ state and (d) the  AFM state.}
 	\label{fig4}
 \end{figure}

Therefore, although $A_5X_3$ is a  band insulator without  freely moving carriers in the bulk, a moderate  electric field along the $c$ axis,  which is strong enough for overcome the ultralow potential barrier, can drive IEs to repeatedly follow the evolution path shown in Figs. \ref{fig3}(a) and (b).
This continuous  transport of IEs will generates a persistent direct current contributed exclusively by the IEs, resulting in the  IE transport proposed here.

The IE transport shares some  similarities with the  topological pumping, where a periodic and finite external perturbation drives the Wannier center of system to traverse the entire unit cell, resulting in quantized charge transport in the insulators. Here, the periodic evolution of IEs along the 1D void channels also constitutes a periodic parameter of the system. The $z$-direction  Wannier center of $A_5X_3$ can be  defined as:
\begin{eqnarray}\label{Zak}
	\Theta_z(k_x,k_y) &=& \int_{0}^{2\pi} {\cal A}_z({\bm k}) d k_z,
\end{eqnarray} 
where ${\cal A}_z$ is the $z$-component of the Berry connection of the system at a given step along the evolution path. As shown in Fig. \ref{fig4}(a), the Wannier centers of both spin-up and spin-down occupied bands change by  a lattice constant along the $c$-axis over one complete  period  shown in Fig \ref{fig3}(a). By treating the periodic evolution  of IEs as a synthetic  dimension, the 3D $A_5X_3$ system can be  mapped onto a (3+1)D topological system with a nontrivial Chern number ${\cal{C}}=2$, where the spin-up and spin-down bands each contribute a topological charge of $1$.

This nontrivial topology can be directly manifested in the boundary states of the system. 
As the IEs undergo a periodic translation along the void channels, two surface states  appear on the (001) surface in succession, and the whole surface bands continuously move across the entire bulk band gap.
The means that under a periodic evolution shown in Fig \ref{fig3}(a),   two electrons in the bulk have been transferred to the (001) surface [see Figs. \ref{fig4}(b-d)].
This gap-crossing evolution of  surface states  not only corroborates the  topological feature of the IE transport, but also provides an intuitive picture for the finite-electric-field induced electronic transport.

\textit{\textcolor{blue}{Discussion.—}}
In this work, we propose a novel electron transport mechanism in electrides, and identify the realistic electride family $A_5X_3$ as candidate materials. The $A_5X_3$ materials are multiferroic materials exhibiting both altermagnetism and type-II FQFE. Distinct from all previously reported  systems, both magnetism and electric polarization in $A_5X_3$ arise solely from the IEs. This unique feature leads to ultralow ferroelectric switching barriers and a novel coupling between FQFE and AM. More importantly, the type-II FQFE feature  of $A_5X_3$ enables the proposal  of TIECs, which exhibit vanishing conductivity at zero temperature but exhibit  persistent electron conduction under a finite static electric field.

Our findings open up several exciting directions for future research. First, the topological IE transport proposed here is ready for  experimental realization, as some $A_5X_3$ family materials have already been experimentally synthesized. Notably, the periodic potential barrier of  IEs in $A_5X_3$ is ultralow. Therefore, the strength of the driving electric field required to achieve IE transport is expected to be well below the dielectric breakdown threshold of the materials. This will allow a clear distinction between topological IE transport and electrical breakdown.

Second, the dependence of TIEC conductivity on external perturbations, such as magnetic field and temperature, should be distinguished from  that in metals, topological insulators, and semiconductors,  representing exciting topics worthy of in-depth investigation in the future.

Finally, beside the $A_5X_3$ family, there exist many other electrides hosting 1D or 2D void channels, such as Mg$_2$Si, Zn$_4$B$_6$O$_{12}$,  Y$_2$Cl$_3$, Gd$_2$Br$_3$, Sr$_5$P$_3$, SnO, PbO, and Sc$_2$C \cite{MizoguchiAngew,liuacsOme,Zhouacs.chemmater, wan_identifying_2018,WangJunjieJACS,Jiaqi024207,doi:10.1021/jacs.2c03024}. Moreover, 2D electrides with delocalized IE layers have been widely predicted and synthesized \cite{Jiaqi024207, doi:10.1021/jacs.2c03024,CHEN2022104176}, suggesting that FQFE and topological IE transport may be generalizable to a broad range of existing electride materials. Searching for new TIECs in electrides and exploring the unique coupling among electric polarization, magnetism, and topological IE transport in these systems will represent another exciting direction for future work.

\acknowledgments

\bibliography{A5X3ref}

\end{document}